\documentclass[pra,superscriptaddress,twocolumn,longbibliography,reprint]{revtex4-2}
\usepackage{amsmath,amsfonts,amssymb,graphics,graphicx,epsfig,color,times,natbib,amsthm,bm,makecell,soul}
\usepackage{booktabs}
\usepackage{dcolumn}
\usepackage{dsfont}
\usepackage{subfigure}
\usepackage{hyperref}
\usepackage{appendix}
\usepackage[ruled]{algorithm2e}

\hypersetup{colorlinks=true, citecolor=blue, urlcolor=blue, linkcolor=blue}
\begin{document}
\date{\today}
\title{Improved Quantum Lattice Boltzmann Method for Advection-Diffusion Equations with a Linear Collision Model}
\author{Li Xu}
\affiliation{College of Science, China University of Petroleum, 266580 Qingdao,  China.}
\author{Ming Li}\email{liming@upc.edu.cn.}
\affiliation{College of Science, China University of Petroleum, 266580 Qingdao,  China.}
\author{Lei Zhang}\email{zhlei84@163.com}
\affiliation{National Key Laboratory of Deep Oil and Gas, China University of Petroleum, 266580 Qingdao, China}
\affiliation{School of Petroleum Engineering, China University of Petroleum, 266580 Qingdao, China}
\author{Hai Sun}
\affiliation{National Key Laboratory of Deep Oil and Gas, China University of Petroleum, 266580 Qingdao, China}
\affiliation{School of Petroleum Engineering, China University of Petroleum, 266580 Qingdao, China}
\author{Jun Yao}\email{RCOGFR_UPC@126.com}
\affiliation{National Key Laboratory of Deep Oil and Gas, China University of Petroleum, 266580 Qingdao, China}
\affiliation{School of Petroleum Engineering, China University of Petroleum, 266580 Qingdao, China}
\begin{abstract}
Quantum computing has made tremendous progress in recent years, providing potentialities for breaking the bottleneck of computing power in the field of scientific computing, like computational fluid dynamics. To reduce
 computational costs and achieve an acceleration, we propose an ancilla free quantum lattice Boltzmann method
 for advection-diffusion equations that fully leverages the parallelism of quantum computing. More significantly, there is no need to perform quantum state tomography in each previous loop, if the macroscopic variables for a certain loop is needed. The non-unitary collision operators are replaced by the unique local unitary operations, and the removal of  ancilla qubit greatly diminishes the complexity of the quantum circuit. The numerical simulations of the $D_1Q_3$ and $D_2Q_5$ models have confirmed the feasibility of the proposed algorithm. 
\end{abstract}

\maketitle
\section{introduction}
Quantum computing has entered the public visual field for the promise of realizing quantum supremacy, by the help of quantum entanglement and quantum superposition \cite{qiqc,nature}. Quantum computing includes quantum computers and quantum algorithms, which complement each other, although the development of the latter is ahead of that of the former \cite{np1,rmp1}. So far, there have been some quantum computers that can solve specific problems beyond the computing power of digital computers, but there is still a lack of generic fault-tolerant quantum computers \cite{qfm,qfm1}. By contrast, the well-known quantum algorithms, such as factorization of a large number via Shor's algorithm \cite{Shor}, the unstructured search using Grover's algorithm \cite{grover}, and the Harrow-Hassidim-Lloyd(HHL) algorithm for solving linear systems \cite{HHL},  which are three prominent examples among many that have been discovered, all rely on fault-tolerant quantum computers to achieve quantum supremacy.
Nevertheless, with the rapid development of quantum computers in recent years, people still firmly believe that fault-tolerant quantum computers can be manufactured in the next few decades \cite{quantum,rmp2}. At that time, the vast majority of quantum algorithms can be implemented, including the quantum lattice Boltzmann method, which requires up to trillions of grid points and millions of time steps for computation, almost exceeding the power of digital computers.

The lattice Boltzmann method(LBM), emerged in the 1990s, is a computational fluid dynamic(CFD) method based on mesoscopic simulation scale. Compared with other CFD methods, LBM has the characteristics of a mesoscopic model that falls between micro molecular dynamics models and macroscopic continuous models, and is widely regarded as an effective means of describing fluid motion and dealing with engineering problems \cite{lbm,benzi,Chen}. 

Recently, efforts are made by Todorova and Steijl for solving the collisionless Boltzmann equation, where the quantum random walk process is taken as a subroutine for executing the streaming step \cite{jcp}. Budinski first proposed the quantum lattice Boltzmann method(QLBM) algorithm for solving advection-diffusion equation, the algorithm introducing an ancilla qubit for the implementation of non-unitary collision operators and the re-normalization of qubit register to support for  multiple successive loops \cite{qip}. However, there exits an inaccuracy in the post-selected segment regarding the macroscopic variables calculation, we will make modifications in Appendix \ref{apa}. Budinski extended the work to Navier-Stokes equations in streamfunction-vorticity formulation \cite{qip2}. Itani and Succi applied the Carleman linearization to the collision term of the lattice Boltzmann formulation, as a first step towards formulating a quantum lattice Boltzmann algorithm \cite{succi1}. Wawrzyniak \textit{et al.} have present a versatile and efficient quantum algorithm based on the LBM from the general three-dimensional case to smaller dimensions and apply to arbitrary lattice-velocity sets \cite{qip3}. In addition, some others have made reformations or summary to LBM \cite{succi,qip4,qip5,qip6}.

We refer to a situation where advection and diffusion occur simultaneously as advection-diffusion process. The one-dimensional advection-diffusion equation can be expressed in Cartesian coordinate system as
\begin{align}
\frac{\partial{\phi}}{\partial{t}}+u\frac{\partial{\phi}}{\partial{x}}=\chi \frac{\partial^2{\phi}}{\partial{x^2}},
\end{align}
where $\phi$ is a dependent variable (mass, momentum, energy, species, etc.),  $u$ is the advection velocity and $\chi$ is the diffusion coefficient. In the manuscript, we adopt periodic boundary conditions for convenience. The finite difference method(FDM) is the most commonly used numerical method to solve the above equation, and the quantum FDM proposed by Berry provides theoretically higher efficiency by the help of HHL algorithm \cite{FDM}. Demirdjian \textit{et al.} have present variational quantum solutions to the advection-diffusion equation, and demonstrated a simulation on the IBM quantum platform \cite{FDM1}. The mathematical structure of LBM is simple and only has a single variable, and it is suitable for parallel computing, which highly coincides with quantum computing.

In this paper, we propose an ancilla free quantum lattice Boltzmann method(AFQLBM) for advection-diffusion equations. The removal of ancilla qubit reduces the complexity of the circuit while preserving the same functionality as the QLBM.  The algorithm can execute any number of loops and does not require quantum state tomography(QST) after each loop,  which is very costly and undoubtedly increases the computational burden. As a replacement of QST, we have proposed a classic algorithm(\textbf{Algorithm \ref{algorithm1}}) specifically for calculating macroscopic variables. In addition, we have designed a unique collision operator for linear equilibrium distribution function---which only requires local unitary operations(LUO) to achieve the goal. The numerical simulations conducted in \textit{qiskit} package not only confirmed the feasibility of the algorithm, but also laid the foundation for executing calculations on real quantum devices in the future.

The remainder of this paper is organized as follows.  In section \ref{section2}, we review the definition of the lattice Boltzmann method.  The theoretical part of the main efforts of this paper is presented in Section \ref{section3}. The numerical simulations are conducted in Section \ref{section4}. The complexity and error analysis are discussed in Section \ref{section5}.  Section \ref{section6} summarizes the main work and future prospects.

\section{Lattice Boltzmann Method}\label{section2}
The single relaxation time lattice Boltzmann equation without source term can be written in the form of
\begin{align}{\label{lbm1}}
f_{\alpha}(x+e_{\alpha}\Delta t,t+\Delta t)-f_{\alpha}(x,t)=\Omega_{\alpha},
\end{align}
where $f_{\alpha}$ is the particle distribution function along the $\alpha$ direction, $x$ is the position defined by the Cartesian coordinate, $e_{\alpha}$ refers to the particle velocity vector and $\Delta t$ is the time step. The Bhatnagar-Gross-Krook(BGK) collision operator
\begin{align}
\Omega_{\alpha}=-\omega[f_{\alpha}(x,t)-f_{\alpha}^{eq}(x,t)],\label{lbm2}
\end{align}
where $\omega = \frac{\Delta t}{\tau}$ and $\tau$ is the single relaxation time.
The local equilibrium distribution function for advection-diffusion equation simply defined as
\begin{align}
f_{\alpha}^{eq}(x,t)=w_{\alpha}\phi(x,t)(1+\frac{e_{\alpha}\cdot\vec{u}}{c_s^2}),
\end{align}
where $w_{\alpha}$ is the  weighting factor in the $\alpha$ direction, $\vec{u}$ is the advection velocity (if $\vec{u}$ is set to be 0, the advection-diffusion problem can be reduced to diffusion problem) and $c_s$ is the speed of sound. By substituting Eq.(\ref{lbm2}) into Eq.(\ref{lbm1}) and arranging it, a more concise form of LBM can be obtained
\begin{align}
f_{\alpha}(x+e_{\alpha}\Delta t,t+\Delta t)=(1-\omega)f_{\alpha}(x,t)+\omega f_{\alpha}^{eq},
\end{align}
 The above equation is the working horse of LBM for the diffusion and advection-diffusion problem. 

The common terminology used in LBM for the dimension of the problem and the number of velocity directions is $D_nQ_m$, where $n=1,2,3$ represents the dimension of the problem and $m$ refers to the number of linkages.  Note that the relation of relaxation time $\tau$ and diffusion coefficient $\chi$ can be deduced by using Chapman-Enskog expansion, which yields 
\begin{align}
\chi=\frac{\Delta x^2}{D_n\Delta t}(\frac{\tau}{\Delta t}-\frac{1}{2}).
\end{align}
Based on the practice situation and for the simplicity of calculation, an appropriate discrete step is chosen such that $\omega=1$.

The solution process LBM mainly consists of two parts, collision and streaming. Particles relax to the local equilibrium condition are executed in the collision step
\begin{align}\label{eq7}
\hat{f}_{\alpha}(x,t)=(1-\omega)f_{\alpha}(x,t)+\omega f_{\alpha}^{eq}.
\end{align}
The following step is to perform streaming operations along each direction, giving that 
\begin{align}\label{eq8}
f_{\alpha}(x+e_{\alpha}\Delta t,t+\Delta t)=\hat{f}_{\alpha}(x,t).
\end{align} 
The macroscopic variable $\phi(x,t)$ is the summing of particle distributions across all directions
\begin{align}\label{eq9}
\phi(x,t)=\sum_{\alpha=0}^{m-1}f_{\alpha}(x,t).
\end{align}
Eq.(\ref{eq7})-Eq.(\ref{eq9}) are a cycle of LBM, through which the macroscopic variables $\phi(x,t)$ can be calculated at a given time $t$.
\section{Ancilla Free Quantum Lattice Boltzmann Method}\label{section3}
\subsection{The $D_1Q_3$ model}
The same as other version of quantum lattice Boltzmann method, the execution of AFQLBM also comprises 4 segments: encoding, collision, streaming, and macroscopic variables calculation. We have improved the steps other than streaming to decrease the quantum resource demanding required for QLBM and fully utilize the parallelism of quantum computing.
(i) Only the initial variables $\phi(x,0)$ need to be encoded, without the need to expand in directions;
(ii) Implementing collision operator through LUO;
(iii) A more convenient method for calculating the macroscopic variables.

AFQLBM only requires two registers (labeled as $q$ and $d$) and does not require the ancilla qubit. Most importantly, the algorithm can achieve any number of loops conditionally controlled by the measurement results in register $q$.

In the encoding step, the initial variable $\phi(x,0)$ in register $q$ is encoded into quantum state through the procedure of Ref.\cite{state}
\begin{align}
|\phi_0(x,0)\rangle=\sum_{i=0}^{M-1}\frac{\phi(i,0)|i\rangle}{\|\phi(x,0)\|_2}.
\end{align}
The number of lattice sites is set to be $M$, $\phi(i,0)$ is the particle mass in lattice site $i$, $|i\rangle$ is the computational basis state, and $\|\cdot\|$ refers to the Euclidean norm. At the beginning of first loop, a crucial quantity $|\phi(x,0)|_1=\sum_i\phi(i,0)$ needs to be recorded because it is necessary when converting the quantum state into macroscopic variables. 

In the collision step, a simpler collision operation can be implemented under the circumstance that the equilibrium distribution $f_{\alpha}^{eq}(x,t)$ and the macroscopic variables $\phi(x,t)$ have a linear relationship(e.g. linear collision model), and these  linear parameters are abbreviated as $\hat{w}_{\alpha}$, i.e. $\hat{w}_{\alpha}=w_{\alpha}(1+\frac{e_{\alpha}\cdot\vec{u}}{c_s^2})$. That is to say,  the distribution function in each direction turns into $\hat{w}_{\alpha}|\psi_0\rangle$ undergoing the collision step. By integrating the states of all directions, we can obtain a normalized state 
\begin{align}
|\phi_1(x,0)\rangle = \frac{\sum_{\alpha=0}^{2}\hat{w}_{\alpha}|\alpha\rangle}{\sqrt{\hat{w}_{0}^2+\hat{w}_{1}^2+\hat{w}_{2}^2}}\otimes |\phi_0\rangle.
\end{align}
To achieve the above local unitary operations on the $d$ register, the $R_y$ gate and controlled-$R_y$ gate with specific rotation angles are indispensable. Appendix \ref{apb} provides the local unitary operations for $D_1Q_3$ model and $D_2Q_5$ model in detail.

Before applying the streaming operator, we reintroduce the $R$ streaming operator \cite{pra}
\begin{align}
R = \sum_{k\in[0,M-1]}|(k+1){\rm{mod}}(M)\rangle\langle k|,
\end{align}
and the $L$ streaming operator 
\begin{align}
L = \sum_{k\in[0,M-1]}|k\rangle\langle(k+1){\rm{mod}}(M)|.
\end{align}
The streaming operators $R$ and $L$ are controlled by the phases of $\hat w_1$ and $\hat w_2$, respectively. After the streaming step, we get 
\begin{align}\label{step3}
|\phi_2(x,0)\rangle &= \frac{\hat w_0|0\rangle|\phi_0\rangle+\hat w_1|1\rangle R|\phi_0\rangle+\hat w_2|2\rangle L|\phi_0\rangle}{\sqrt{\hat{w}_{0}^2+\hat{w}_{1}^2+\hat{w}_{2}^2}}\notag\\
&=\frac{|0\rangle f_0(x,1)+|1\rangle f_1(x,1)+|2\rangle f_2(x,1)}{\sqrt{\hat{w}_{0}^2+\hat{w}_{1}^2+\hat{w}_{2}^2}}.
\end{align}
$f_i(x,1)$ is the state after the flow in the $i$ direction ends, $i=0,1,2$.
\begin{figure}[t]
\centering\includegraphics[width=0.5\textwidth]{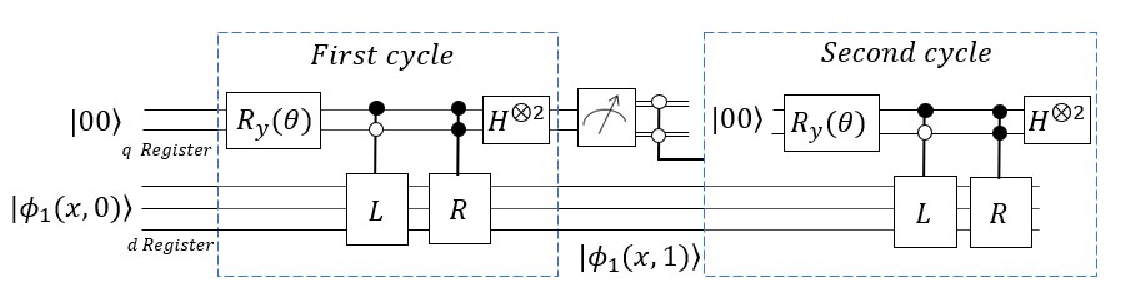}
\caption{The circuit diagram of $D_1Q_3$ model for modified quantum lattice Boltzmann method.}\label{Fig1}
\end{figure}

\begin{algorithm}[b]\label{algorithm1}
\caption{Macroscopic variables calculation}
\LinesNumbered
\emph{Input.} The number $S$ of shots and $|\phi(x,0)|_1$.\\
\emph{Output.} New macroscopic variables $\phi(x,1)$.\\
\emph{1.} Record the number of measurements for each phase $S_i,i=0,1,...,M-1$.\\
\emph{2.} Calculate $\sqrt{S_i}/\sqrt{S}$ and the summation $S_{all}$ of all $\sqrt{S_i}/\sqrt{S}$.\\
\emph{3.} Calculate the weight of each phase $\sqrt{S_i}/(\sqrt{S}\cdot S_{all})$.\\
\emph{4.} Assign $|\phi(x,0)|_1$ to each phase according the weight to obtain $\phi(x,1)$.
\end{algorithm}

In the final step, the calculation of the macroscopic variables by point-wise addition of the right-hand side of Eq.(\ref{step3}) is performed. This summation process is simply achieved by applying Hadamard gates to the $q$ register
\begin{align}
|\phi_3(x,0)\rangle =& (H_q^{\otimes 2}\otimes I_d)|\phi_2\rangle\notag\\
=&\frac{|00\rangle[f_0(x,1)+f_1(x,1)+f_2(x,1)]}{2\sqrt{\hat{w}_{0}^2+\hat{w}_{1}^2+\hat{w}_{2}^2}}\notag\\
&+\frac{|01\rangle|\cdot\rangle+|10\rangle|\cdot\rangle+|11\rangle|\cdot\rangle}{2\sqrt{\hat{w}_{0}^2+\hat{w}_{1}^2+\hat{w}_{2}^2}},
\end{align}
where the subscripts of $H$ and $I$ represent the registers the quantum gates applied to, and $|\cdot\rangle$ refers to the trivial combinations of $f_0(x,1)$, $f_1(x,1)$ and $f_2(x,1)$. If measurements are taken on the $q$ register, the collapsed quantum state $|\phi_0(x,1)\rangle=N_1[f_0(x,1)+f_1(x,1)+f_2(x,1)]$ can be obtained, on the premise that the results are $|00\rangle$. $N_1$ is a normalized factor for the first loop. Appendix \ref{apc} discusses in detail the probability of the collapse of the  register $q$.

The good news is that $|\phi_0(x,1)\rangle$ can serve as the initial state if the loop does not terminate. If the $\phi(x,1)$ need to be extracted, we just take measurements to the $d$ register and then use $|\phi(x,0)|_1$ to calculate new macroscopic variables. We provide the calculation process for macroscopic variables in \textbf{Algorithm \ref{algorithm1}} and the entire algorithm flowchart in Fig.\ref{Fig1}.

\subsection{The $D_2Q_5$ model}
Most cases of $D_2Q_5$ model keep consistent with $D_1Q_3$ model except for the streaming step,  and the increase in dimensionality rises complexity of the streaming operations---two particle velocity directions along the $y$-axis are added.

The same encoding strategy is employed for the preparation of initial state $|\phi_0(x,0)\rangle$, and the dimension of the initial state is equal to the number of lattice cells. We need 3 qubits to execute the linear collision in the $q$ register, and the detailed operators are shown in Appendix \ref{apb}. After the collision step, the state evolves into
\begin{align}
|\phi_1(x,0)\rangle = \frac{\sum_{\alpha=0}^{4}\hat{w}_{\alpha}|\alpha\rangle}{\sqrt{\hat{w}_{0}^2+\hat{w}_{1}^2+\hat{w}_{2}^2+\hat{w}_{3}^2+\hat{w}_{4}^2}}\otimes |\phi_0\rangle.
\end{align}

In the streaming step, apart from the particles remaining at the origin, there are a total of 4 streaming directions, 2 along the $x$-axis and 2 along the $y$-axis. The streaming operators for the 4 directions are as follows
\begin{align}
S_1 = R\otimes I_M, S_2 = L\otimes I_M,\notag\\
S_3 = I_M\otimes R, S_4 = I_M\otimes L.
\end{align}
And these operators are controlled by the phases of $\hat{w}_{\alpha}$, $\alpha=1,2,3,4$. The state after streaming  is
\begin{align}
|\phi_2(x,0)\rangle &= \frac{\hat w_0|0\rangle|\phi_0\rangle+\sum_{\alpha=1}^4\hat w_{\alpha}|\alpha\rangle S_{\alpha}|\phi_0\rangle}{\sqrt{\hat{w}_{0}^2+\hat{w}_{1}^2+\hat{w}_{2}^2+\hat{w}_{3}^2+\hat{w}_{4}^2}}\notag\\
&=\frac{\sum_{\alpha=0}^4f_{\alpha}(x,1)}{\sqrt{\hat{w}_{0}^2+\hat{w}_{1}^2+\hat{w}_{2}^2+\hat{w}_{3}^2+\hat{w}_{4}^2}}.
\end{align}

\begin{figure}[t] 
\centering\includegraphics[width=0.5\textwidth]{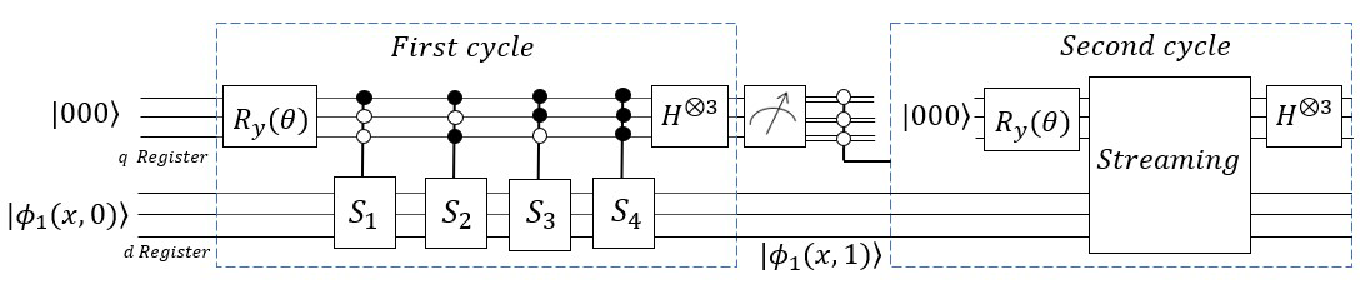}
\caption{The circuit diagram of $D_2Q_5$ model for modified quantum lattice Boltzmann method.}\label{Fig2}
\end{figure}

 The calculation of macroscopic variables is similarly consistent with $D_1Q_3$ model, the only difference is that an additional $H$ gate is required because the $q$ register has 3 qubits. The circuit diagram is shown in FIG. \ref{Fig2}.

\section{Numerical Simulation}\label{section4}
Numerical simulations for $D_1Q_3$ and $D_2Q_5$ models are conducted on the \textit{qasm\_simulator} backend in the \textit{qiskit} package to verify the feasibility of the proposed algorithm. Both models adopt periodic boundaries, and the measurement frequency is selected as the number of lattice cells multiplied by $10^4$. The issue that must be mentioned is that we use $\Delta \phi$ instead of directly using $\phi$ for calculation in the numerical simulation process, in order to obtain more accurate calculation results. We profoundly investigate this issue when conducting the complexity and feasibility analysis of the algorithm in the next section.
\subsection{The $D_1Q_3$ model}
\begin{figure}[b]
\centering\includegraphics[width=0.5\textwidth]{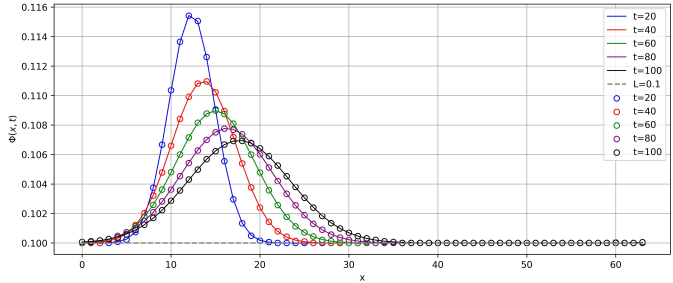}
\caption{Numerical comparison results of the quantum (`o') and classical (`---') LBM. }\label{Fig3}
\end{figure}
In this simulation, movements of Gaussian hill are conducted on the $D_1Q_3$ model. The settings for each parameter are as follows: $\Delta x = 1$, $\Delta t = 1$, $u = 0.2$, $\omega=1$ and $\chi=0.5$, all in lattice units. There are 64 lattice cells in total, and the initial variables $\phi(x,0)=0.1$ at each lattice cell except for $\phi(11,0)=0.2$. The numerical results shown in FIG. \ref{Fig3} reveal that AFQLBM and LBM have very good consistency.

\subsection{The $D_2Q_5$ model}
\begin{figure}[t]
\subfigure[]{\includegraphics[width=0.23\textwidth]{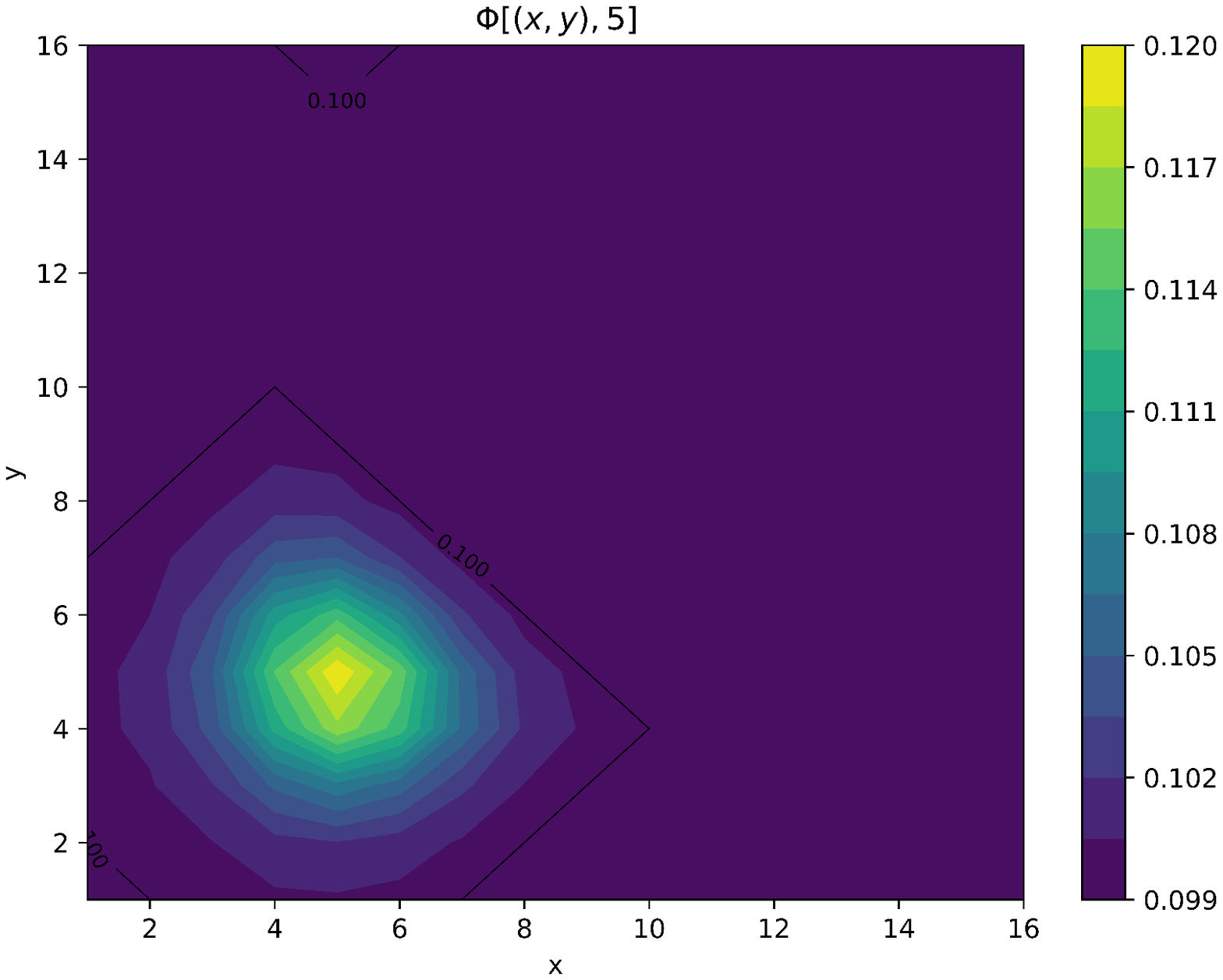}\label{fig4a}}
\subfigure[]{\includegraphics[width=0.23\textwidth]{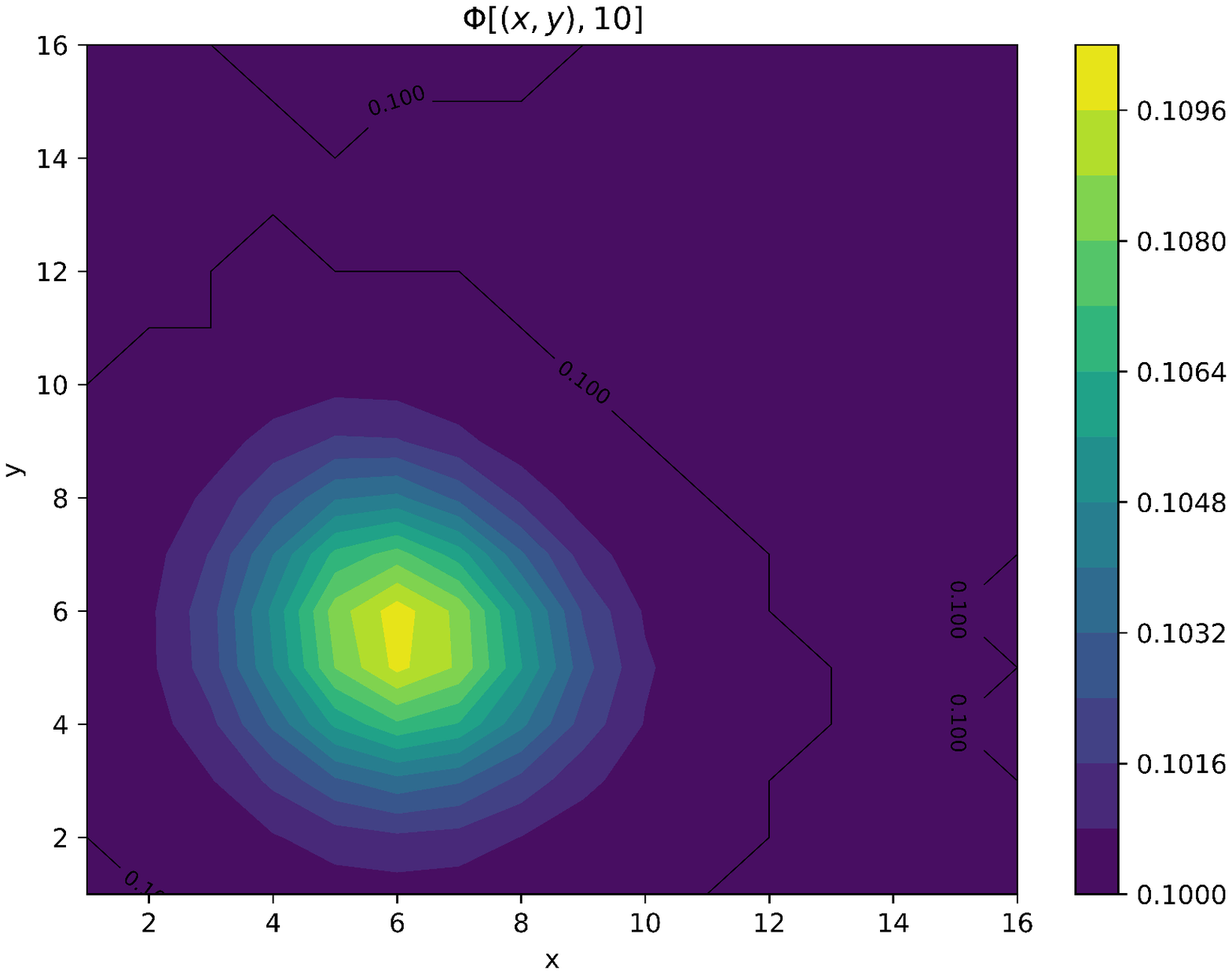}\label{fig4b}}\\
\subfigure[]{\includegraphics[width=0.23\textwidth]{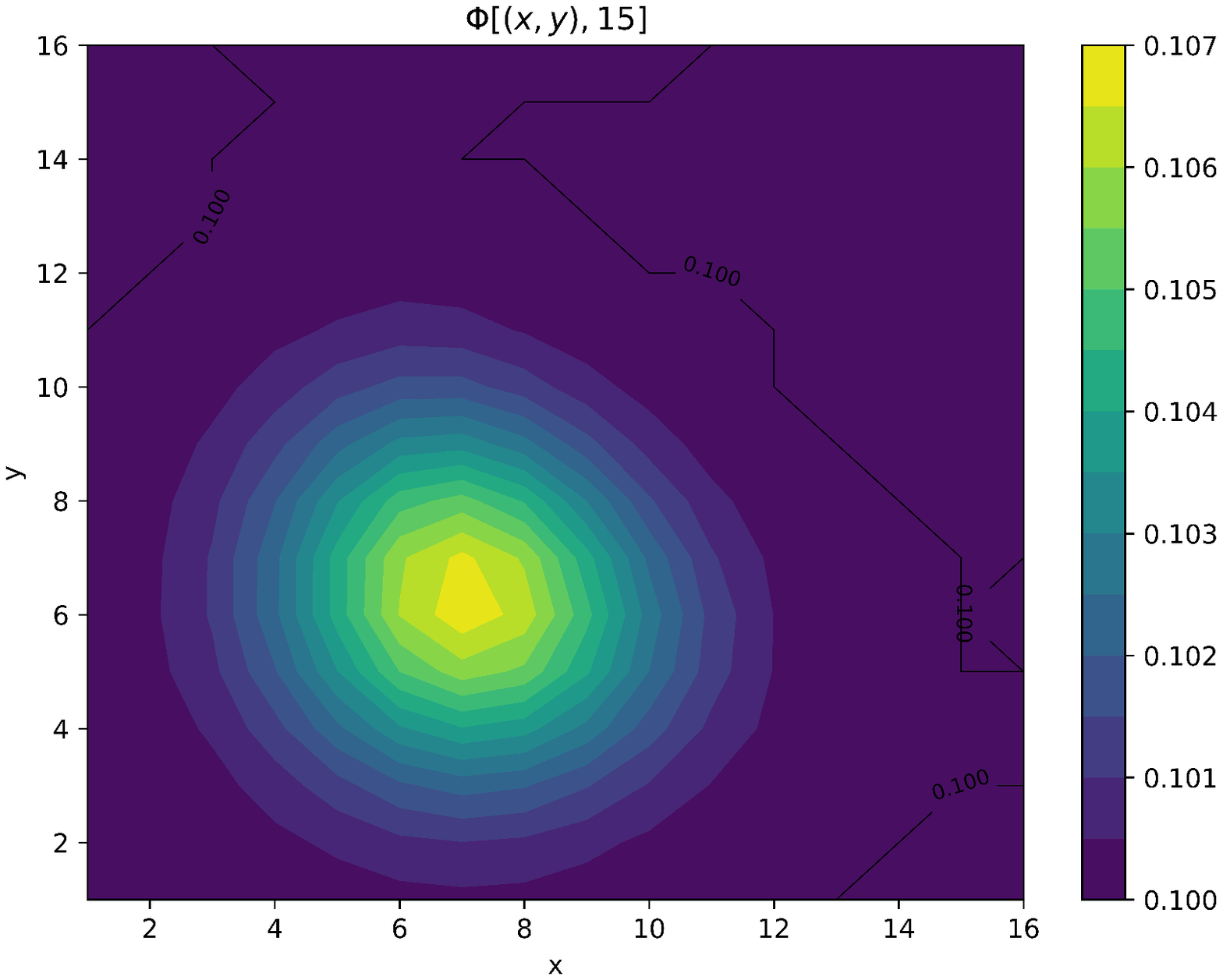}\label{fig4c}}
\subfigure[]{\includegraphics[width=0.23\textwidth]{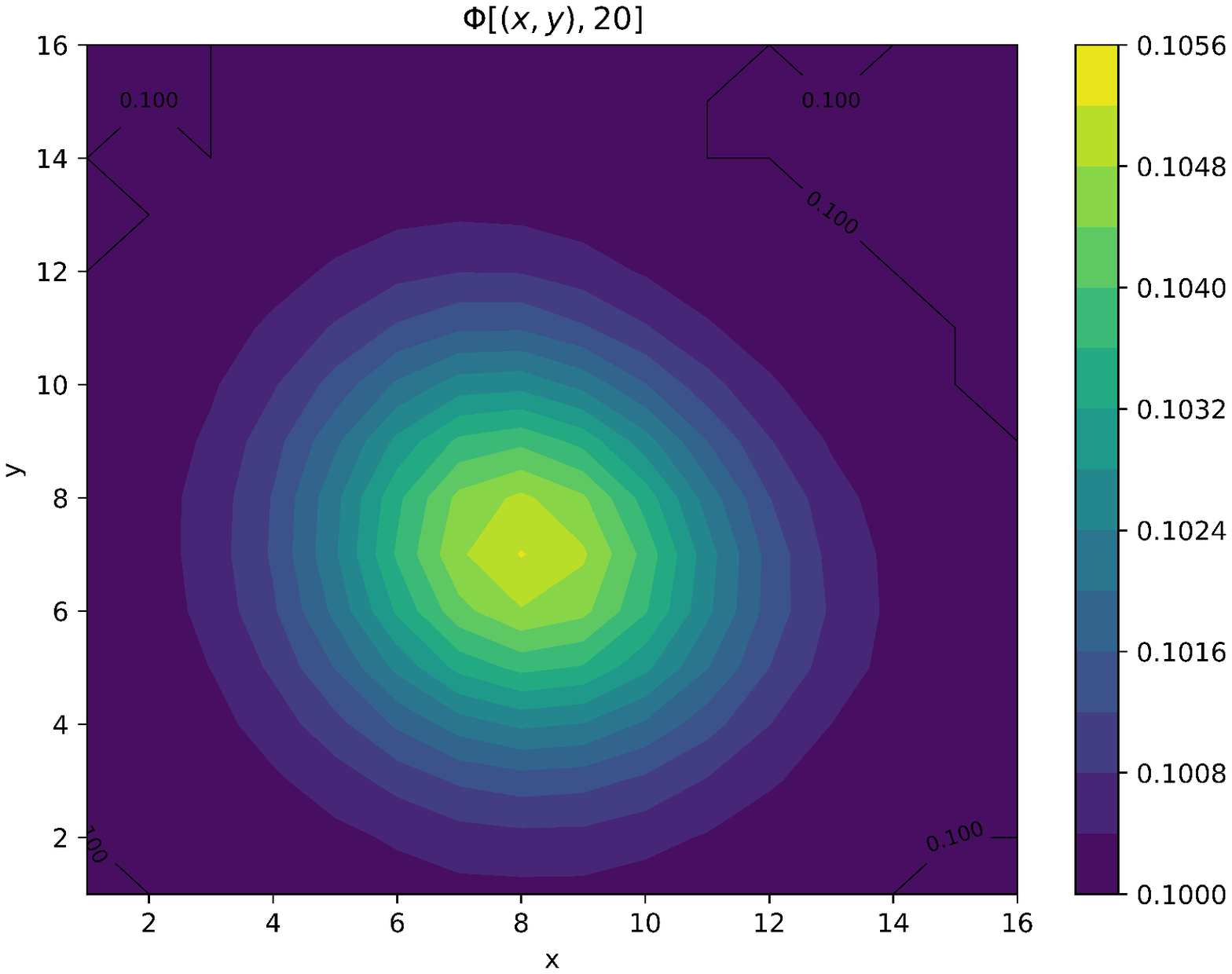}\label{fig4d}}
\caption{Numerical results running on the $D_2Q_5$ model with time interval $t = 5$.}\label{Fig4}
\end{figure}

The $D_2Q_5$ model for Gaussian hill is conducted on a $16\times 16$ square lattice layout. The settings for each parameter are as follows: $\Delta x = 1$, $\Delta y = 1$, $\Delta t = 1$, $u = 0.2$, $v = 0.15$, $\omega=1$ and $\chi=1/6$, all in lattice units. The initial variables $\phi[(x,y),0]=0.1$ at each lattice cell except for $\phi[(4,4),0]=0.3$. The numerical results shown in FIG. \ref{Fig4} showcase the motion process of Gaussian hill.

\section{Complexity analysis and error analysis}\label{section5}
The streaming steps are the dominant factor in the complexity of AFQLBM, the principal reasons are that the state preparation is only required in the first loop and the local unitary operators just occupy a few qubits(no more than 5). The execution of streaming step demands a series of multiple controlled-NOT gates.  In this paper, the count number of Toffoli gates is referred as the benchmark of the complexity of the circuit.

Take a $D_2Q_5$ model with $M\times M$ lattice cells as an example. There are 3 qubits in the $q$ register and $2\log_2{M}$ qubits in the $d$ register. It should be noted that an $n$-controlled-NOT gate can be decomposed into $2n-3$ Toffoli gates. There are 4 directions that require the streaming operators. The number of controlled qubits for the multiple controlled-NOT gates range from 3 to $\log_2{M}+2$ in each direction, converted to Toffoli gates count $3,5,7,...,2\log_2{M}+1$. The conclusion that the number of Toffoli gates required in one loop is $4\log_2^2{M}+8\log_2{M}$. Compared to the previous method in \cite{qip,qip3} which needs $4\log_2^2{M}+16\log_2{M}$ Toffoli gates, there is a reduction in the magnitude of $O(\log_2 M)$. For the current stage of real quantum devices and quantum simulators, this is a significant improvement.

\begin{figure}[t]
\includegraphics[width=0.5\textwidth]{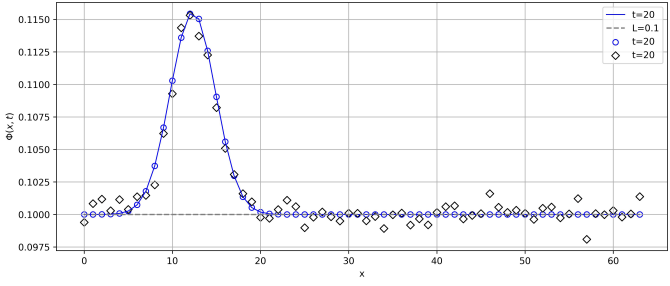}\label{fig5}
\caption{Numerical comparison results for the difference method: `---' is the classical results, `o' is the quantum results obtained using the difference method, and `$\diamond$' is the results obtained using direct encode method.}\label{Fig5}
\end{figure}

The sources of error are device noise and finite sampling noise, we conduct a simple analysis of the latter and propose a trick to mitigate this error. Reconstructing a full description of a quantum system drawing support from quantum state tomography necessitates a number of measurement repetitions exponential in qubit number \cite{np}. This will lead to falling into the curse of dimensionality. Regarding our issue, special methods can be adopted to reduce the number of measurements while minimizing read-out errors. When taking measurements to a quantum state, it collapses to the phase with the probability of the corresponding amplitude squared. The information of quantum state is extracted through the collapsing number of measurements for each phase. Macroscopic variables are stored in the amplitude of quantum states. If the difference method is used, a large part of the phase amplitude will become 0, which undoubtedly greatly reduces the difficulty of read-out. We conduct comparative experiments on the $D_1Q_3$ model(FIG. \ref{Fig5}) and the results reveal that the error of the difference method is significantly smaller.

\section{Conclusion and Prospect}\label{section6}
The main contribution of this manuscript is to devise an ancilla free quantum algorithm of the lattice Boltzmann method for solving advection-diffusion equation. The removal of ancilla qubit reduces the complexity of the circuit while permitting the algorithm to loop conditionally controlled by the partial measurement results. In theory, macroscopic variables can be calculated at any time step, as long as our device allows for the calculation of deep circuits.  Furthermore, the calculation of the current macroscopic variables only requires the 1-norm of the initial variables and the read out results of the current loop, without the need of reading out the value of the previous cycles. If the current iterative results of the algorithm need to read out the previous calculation results, it is not advisable for quantum computing. The reason is that the conversion of quantum system information to classical information is accompanied by errors, which gradually increase in the cyclic process, thus undermine the accuracy of the final results. And AFQLBM precisely avoids this obstacle that iterative algorithms often encounter, so it has better application prospects. AFQLBM is not the quantization of classical LBM, but a practical quantum algorithm that can be implemented on real quantum computers.

The follow-up work needs to incorporate more complex boundary conditions into AFQLBM and consider more practical gradient advection velocities. The proposed algorithm may be expanded to incompressible Navier-Stokes equation and multiphase flow model.

\section*{Acknowledgments} This work is supported by the Shandong Provincial Natural Science Foundation for Quantum Science No. ZR2021LLZ002,  the Fundamental Research Funds for the Central Universities No. 22CX03005A, and The National Natural Science Foundation of China (Grant Nos. 52122402, and 52034010).

\begin{appendix}
\section{Reconstruction of Ref.\cite{qip}{\label{apa}}}
We demonstrate a proof-of-principle to point out the matters that Ref.\cite{qip} owns. The parameters are setting to be $\Delta x = 1$, $\Delta t = 1$, $u = 0.5$, $\omega=1$ and $\chi=0.5$. The initial macroscopic variables are $
\phi(x,0)=(0,0,1,0)$. We use the $D_1Q_2$ model, and $\hat w_1=0.75$, $\hat w_2=0.25$. Obviously, the macroscopic variables for the next moment are $\phi(x,1)=(0,0.25,0,0.75)$.

Next, we will strictly follow the steps in Ref.\cite{qip} to calculate the evolution process of the state. We have $\Phi^0=(0,0,1,0,0,0,1,0)^T$, and the $\Phi$ can be encoded into quantum state as
\begin{align}
|\phi_0\rangle =& |0\rangle_a\frac{1}{\sqrt 2}(|001\rangle+|101\rangle),
\end{align}
the norm of $\|\Phi^0\|=\sqrt 2$. The diagonal matrix$A = {\rm{diag}}(0.75,0.75,0.75,0.75,0.25,0.25,0.25,0.25)$. 
The cillision operator 
\begin{align}
(H\otimes I)(|0\rangle\langle 0|_a\otimes C_1+|1\rangle\langle 1|_a\otimes C_2)(H\otimes I)
\end{align}
where $C_1=A+i\sqrt{I-A^2}$, $C_2=A-i\sqrt{I-A^2}$. Following the collision step, the initial quantum state $|\phi_0\rangle$ has evolved into
\begin{align}
|\phi_1\rangle =&(H\otimes I)(|0\rangle\langle 0|_a\otimes C_1+|1\rangle\langle 1|_a\otimes C_2)(H\otimes I)|\phi_0\rangle\\
=&(H\otimes I)(|0\rangle\langle 0|_a\otimes C_1+|1\rangle\langle 1|_a\otimes C_2)\notag\\
&\times\frac{1}{2}(|0\rangle_a+|1\rangle_a)(|001\rangle+|101\rangle)\\
=&(H\otimes I)\frac{1}{2}[|0\rangle_aC_1(|001\rangle+|101\rangle)\notag\\
&+|1\rangle_aC_2(|001\rangle+|101\rangle)]\\
=&\frac{1}{2\sqrt{2}}[(|0\rangle_a+|1\rangle_a)C_1(|001\rangle+|101\rangle)\notag\\
&+(|0\rangle_a-|1\rangle_a)C_2(|001\rangle+|101\rangle)]\\
=&\frac{1}{2\sqrt{2}}[|0\rangle_a(C_1+C_2)(|001\rangle+|101\rangle)\notag\\
&+|1\rangle_a(C_1-C_2)(|001\rangle+|101\rangle)]\\
=&\frac{1}{\sqrt{2}}|0\rangle_aA(|001\rangle+|101\rangle)\notag\\
&+\frac{1}{\sqrt{2}}|1\rangle_ai\sqrt{I-A^2}(|001\rangle+|101\rangle).
\end{align}
In the streaming step, the quantum state of the entire system is
\begin{align}
|\phi_2\rangle  =& \frac{1}{\sqrt{2}}|0\rangle_aRLA(|001\rangle+|101\rangle)\notag\\
&+\frac{1}{\sqrt{2}}|1\rangle_ai\sqrt{I-A^2}(|001\rangle+|101\rangle)\\
=&\frac{1}{\sqrt{2}}|0\rangle_aRL(0.75|001\rangle+0.25|101\rangle)\notag\\
&+\frac{1}{\sqrt{2}}|1\rangle_a(\frac{i\sqrt 7}{4}|001\rangle+\frac{i\sqrt{15}}{4}|101\rangle)\\
=&\frac{1}{\sqrt{2}}|0\rangle_a(0.75|000\rangle+0.25|110\rangle)\notag\\
&+\frac{1}{\sqrt{2}}|1\rangle_a(\frac{i\sqrt 7}{4}|001\rangle+\frac{i\sqrt{15}}{4}|101\rangle).
\end{align}
In the last step, the calculation of the macroscopic variables are calculated after applying the SWAP gate and Hadamard gate
\begin{align}
|\phi_3\rangle =&(H_a\otimes I){\rm{SWAP}}[\frac{1}{\sqrt{2}}|0\rangle_a(0.75|000\rangle\notag\\
&+0.25|110\rangle)+\frac{1}{\sqrt{2}}|1\rangle_ai\sqrt{I-A^2}(|001\rangle+|101\rangle)]\\
=&(H_a\otimes I){\rm{SWAP}}[\frac{0.75}{\sqrt{2}}|0\rangle_a|000\rangle+\frac{0.25}{\sqrt{2}}|0\rangle_a|110\rangle\notag\\
&+\frac{i\sqrt 7}{4\sqrt{2}}|1\rangle_a|001\rangle+\frac{i\sqrt{15}}{4\sqrt{2}}|1\rangle_a|101\rangle]\\
=& (H_a\otimes I)[\frac{0.75}{\sqrt{2}}|0\rangle_a|000\rangle+\frac{0.25}{\sqrt{2}}|1\rangle_a|010\rangle\notag\\
&+\frac{i\sqrt 7}{4\sqrt{2}}|0\rangle_a|101\rangle+\frac{i\sqrt{15}}{4\sqrt{2}}|1\rangle_a|101\rangle]\\
=&\frac{0.75}{2}(|0\rangle_a+|1\rangle_a)|000\rangle+\frac{0.25}{2}(|0\rangle_a-|1\rangle_a)|010\rangle\notag\\
&+\frac{i\sqrt 7}{8}(|0\rangle_a+|1\rangle_a)|101\rangle+\frac{i\sqrt{15}}{8}(|0\rangle_a-|1\rangle_a)|101\rangle\\
=&|0\rangle_a(\frac{0.75}{2}|000\rangle+\frac{0.25}{2}|010\rangle+\frac{i(\sqrt 7 +\sqrt{15})}{8}|101\rangle)\notag\\
&+|1\rangle_a(\frac{0.75}{2}|000\rangle-\frac{0.25}{2}|010\rangle+\frac{i(\sqrt 7 -\sqrt{15})}{8}|101\rangle).
\end{align}
By post-selecting the ancilla qubit in $|0\rangle_a$ and mutiplying the factor $\frac{2\|\phi\|}{\sqrt 2}=2$, we get 
$|\phi_4\rangle = 0.75|000\rangle+0.25|010\rangle+\frac{i(\sqrt 7 +\sqrt{15})}{4}|101\rangle$. The author of Ref.\cite{qip} claimed that the spatial distribution of the variable $\phi$ for the next time level $t+1$ is obtained, and the entire procedure is then repeated to achieve desired time level. However,  although $\phi(x,1)$ can be extracted from $|\phi_4\rangle$, under the premise that the first qubut is $0$. That is to say, the previous calculation result not only requires post-selected of ancilla qubit, but also includes additional qubits. On the other hand, at the end of each time step, quantum state tomography is necessary because the modulus at time $t$ is needed to calculate the distribution function at time $t+1$. Quantum state tomography will cause the curse of dimensionality, and quantum computing should avoid QST as much as possible, especially cyclic algorithms.
\begin{figure}[t]
\subfigure[]{\includegraphics[width=0.15\textwidth]{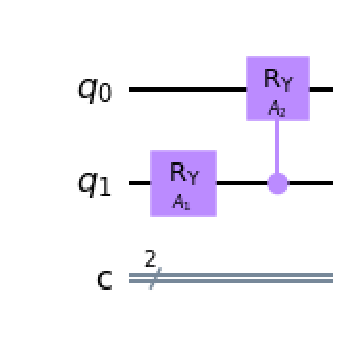}\label{figa1}}
\subfigure[]{\includegraphics[width=0.3\textwidth]{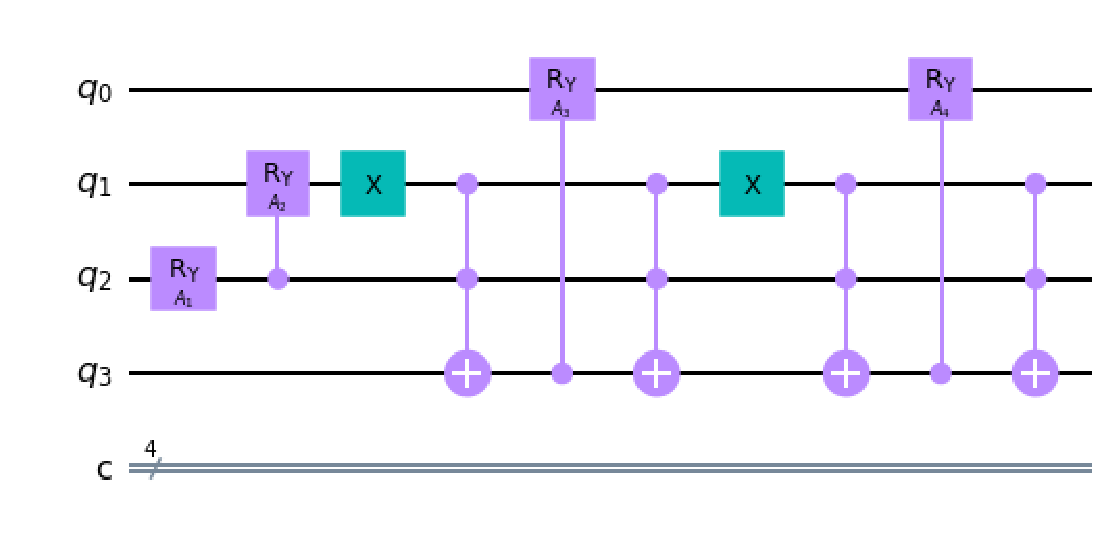}\label{figa2}}\\
\caption{Local unitary operations for $D_1Q_3$ and $D_2Q_5$ model.  $q_3$ is a work qubit.}\label{Figa1}
\end{figure}
\section{Local Unitary Operations{\label{apb}}}
In this section, local unitary operations for the collision step for $D_1Q_3$ and $D_2Q_5$ are discussed in detail. The number of qubits required for this encoding process equals to $\lceil\log_2{m}\rceil$. That is to say, the two models require 2 and 3 qubits respectively. 

For $D_1Q_3$ model, the rotations angles are calculated through 
\begin{align}
A_1 &= 2\arccos{\frac{\hat{w}_{0}}{\sqrt{\hat{w}_0^2+\hat{w}_1^2+\hat{w}_2^2}}},\\
A_2 &= 2\arccos{\frac{\hat{w}_{1}}{\sqrt{\hat{w}_1^2+\hat{w}_2^2}}},
\end{align}
where  $\hat{w}_{\alpha}$ is consistent with the main text, $\hat{w}_{\alpha}=w_{\alpha}(1+\frac{e_{\alpha}\cdot\vec{u}}{c_s^2})$. Then the collision operator can be constructed
\begin{align}
[C_1\textendash R_y(A_2)]\times[R_y(A_1)\otimes I_2],
\end{align}
where $C_1\textendash R_y(A_2)$ represents the controlled$\textendash R_y$ gate and the subscript of $C$ means the gate is controlled by qubit 1.

Applying the collision operator to initial state $|00\rangle$, one gets the normalized state
\begin{align}
\frac{\hat{w}_{0}|00\rangle+\hat{w}_{1}|10\rangle+\hat{w}_{2}|11\rangle}{\sqrt{\hat{w}_0^2+\hat{w}_1^2+\hat{w}_2^2}},
\end{align}
and the subsequent streaming step in each direction is controlled by the phase of $\hat{w}_{\alpha}$, i.e. $|10\rangle$ controlls direction $1$ and $|11\rangle$ controlls direction $2$.

For $D_2Q_5$ model, the rotations angles are calculated through 
\begin{align}
A_1 &= 2\arccos{\frac{\hat{w}_{0}}{\sqrt{\hat{w}_0^2+\hat{w}_1^2+\hat{w}_2^2+\hat{w}_3^2+\hat{w}_4^2}}},\\
A_2 &= 2\arccos{\frac{\sqrt{\hat{w}_1^2+\hat{w}_2^2}}{\sqrt{\hat{w}_1^2+\hat{w}_2^2+\hat{w}_3^2+\hat{w}_4^2}}},\\
A_3 &= 2\arccos{\frac{\hat{w}_1}{\sqrt{\hat{w}_1^2+\hat{w}_2^2}}},\\
A_4 &= 2\arccos{\frac{\hat{w}_3}{\sqrt{\hat{w}_3^2+\hat{w}_4^2}}}.
\end{align}
Then the collision operator has the following form
\begin{align}
&[C_1C_0\textendash R_y(A_4)]\times[C_1C_1\textendash R_y(A_3)]\times\notag\\
&[C_1\textendash R_y(A_2)\otimes I_2]\times[R_y(A_1)\otimes I_4],
\end{align}
where the $CC\textendash R_y$ gate means that the $R_y$ gate has two controlled qubits. The initial state $|000\rangle$ undergoes the collision operation turns to be 
\begin{align}
\frac{\hat{w}_{0}|000\rangle+\hat{w}_{1}|100\rangle+\hat{w}_{2}|101\rangle+\hat{w}_{3}|110\rangle+\hat{w}_{4}|111\rangle}{\sqrt{\hat{w}_0^2+\hat{w}_1^2+\hat{w}_2^2+\hat{w}_3^2+\hat{w}_4^2}}.
\end{align}
The subsequent streaming operations are controlled by the phases of $\hat{w}_{\alpha}$ like the $D_1Q_3$ model. We have provided the quantum circuit(FIG. \ref{Figa1}) for implementing local unitary operations for both models in \textit{qiskit} package.

\section{Probability Analysis{\label{apc}}}
In this section, we will discuss the probability of register $q$ collapsing to 0 string through measurement. Before taking measurement on the $q$ register, we have the following state
\begin{align}\label{C1}
|\phi_3(x,0)\rangle =&\frac{|00\rangle[f_0(x,1)+f_1(x,1)+f_2(x,1)]}{2\sqrt{\hat{w}_{0}^2+\hat{w}_{1}^2+\hat{w}_{2}^2}}\notag\\
&+\frac{|01\rangle|\cdot\rangle+|10\rangle|\cdot\rangle+|11\rangle|\cdot\rangle}{2\sqrt{\hat{w}_{0}^2+\hat{w}_{1}^2+\hat{w}_{2}^2}}.
\end{align}
For the sake of simplicity and convenience, we use $f_i$ instead of $f_i(x,1)$, $i=0,1,2$. From Eq.(\ref{step3}), we have
\begin{align}
\frac{f_0^2+f_1^2+f_2^2}{\hat{w}_{0}^2+\hat{w}_{1}^2+\hat{w}_{2}^2}=1.
\end{align} 
To estimate the probability of register $q$ collapsing to $|00\rangle$ in Eq.(\ref{C1}),  we need to calculate the interval where the square of its amplitude is located, i.e.,
\begin{align}
\frac{(f_0+f_1+f_2)^2}{4(\hat{w}_{0}^2+\hat{w}_{1}^2+\hat{w}_{2}^2)}.
\end{align}
 Without loss of generality, we assume that $\hat{w}_{0}^2+\hat{w}_{1}^2+\hat{w}_{2}^2=1$. The problem becomes that given 3 non-negative number $f_0$, $f_1$, $f_2$, and satisfy $f_0^2+f_1^2+f_2^2=1$, find the range of values for $\frac{(f_0+f_1+f_2)^2}{4}$. According to Cauchy's inequality, we have
\begin{align}
f_0+f_1+f_2\leq \sqrt{3(f_0^2+f_1^2+f_2^2)}=\sqrt 3.
\end{align}
Obviously, we have $(f_0+f_1+f_2)\geq 1$, giving that 
\begin{align}
\frac{1}{4}\leq \frac{(f_0+f_1+f_2)^2}{4}\leq \frac{3}{4}.
\end{align}

Hence we can conclude that the probability locates in $[\frac{1}{4},\frac{3}{4}]$. The above is a rigorous mathematical derivation, and when we combine it with practical situations, we will find that some situations may not occur. For example, when the probability is $\frac{1}{4}$, one of $f_0$, $f_1$, and $f_2$ needs to be 1, and the other two need to be equal to 0. This contradicts the advection-diffusion phenomenon. When the probability is $\frac{3}{4}$, $f_0=f_1=f_2=\frac{\sqrt 3}{3}$ is required. This means that the particles will be uniformly distributed in the adjacent 3 lattices at the next moment, which is also contrary to the advection-diffusion phenomenon.

This probability is related to the diffusion coefficient and advection velocity, and the probability is 72\% in our numerical simulation. By adjusting different values, we find that this probability is generally not less than 50\%.

\end{appendix}


\begin{thebibliography}{99}
\bibitem{qiqc} M. A. Nielsen and I. Chuang. \textit{Quantum Computation and Quantum Information} (Cambridge University Press, Cambridge, 2002).
\bibitem{nature}  F. Arute, K. Arya, R. Babbush, \textit{et al.}, Quantum supremacy using a programmable superconducting processor. \href{ https://doi.org/10.1038/s41586-019-1666-5}{Nature 574, 505-510 (2019)}.
\bibitem{rmp1}  K. Bharti,  A. Cervera-Lierta, T. H. Kyaw, M. Degroote, H. Heimonen,  J. S. Kottmann,  \textit{et al.},  Noisy intermediate-scale quantum algorithms. \href{https://journals.aps.org/rmp/abstract/10.1103/RevModPhys.94.015004}{REVIEWS OF MODERN PHYSICS 94 (2022)}.
\bibitem{np1} S. Boixo,   S. V. Isakov,  V. N. Smelyanskiy,  \textit{et al.}, Characterizing quantum supremacy in near-term devices. \href{https://doi.org/10.1038/s41567-018-0124-x}{Nature Physics 14, 595–600 (2018)}. 
\bibitem{qfm}  M. Schuld and N. Killoran, Quantum machine learning in feature Hilbert spaces, \href{https://doi.org/10.1103/PhysRevLett.122.040504}{Phys. Rev. Lett. 122, 040504 (2019)}.
\bibitem{qfm1}  V. Havl\'{i}$\rm\check{c}$ek, A. D. C\'{o}rcoles,  K. Temme,  A. W. Harrow, A. Kandala, J. M. Chow, and J. M. Gambetta, Supervised learning with quantum-enhanced feature spaces, \href{https://doi.org/10.1038/s41586-019-0980-2}{Nature 567, 209-212 (2019)}. 
\bibitem{Shor} P. Shor, Polynomial-Time Algorithms for Prime Factorization and Discrete Logarithms on a Quantum Computer, \href{https://epubs.siam.org/doi/10.1137/S0036144598347011}{SIAM Journal on Computing, 26, 1484(1997)}.
\bibitem{grover} L. K. Grover, Quantum Mechanics Helps in Searching for a Needle in a Haystack, \href{https://doi.org/10.1103/PhysRevLett.79.325}{Phys. Rev. Lett. 79, 325 (1997)}.
\bibitem{HHL}A. W. Harrow , A. Hassidim,  and S. Lloyd,  Quantum algorithm for linear systems of equations, \href{https://doi.org/10.1103/PhysRevLett.103.150502}{Phys. Rev. Lett. 103, 150502 (2009)}.
\bibitem{quantum} P. John, Quantum Computing in the NISQ era and beyond. \href{	https://doi.org/10.22331/q-2018-08-06-79}{Quantum 2,79 (2018)}.
\bibitem{rmp2} C.-Y. Lu, Y. Cao, C.-Z. Peng, and J.-W. Pan, Micius quantum experiments in space. \href{https://journals.aps.org/rmp/abstract/10.1103/RevModPhys.94.035001}{REVIEWS OF MODERN PHYSICS 94 (2022)}.
\bibitem{lbm} A. A. Mohamad, \textit{Lattice Boltzmann Method}(Fundamentals and Engineering Applications with Computer Codes).
\bibitem{benzi} R. Benzi, S. Succi, M. Vergassola, The lattice Boltzmann equation: theory and applications, \href{https://doi.org/10.1016/0370-1573(92)90090-M}{Phys. Rep. 222, 145(1992)}.
\bibitem{Chen} S. Chen, and D. D. Doolen, Lattice Boltzmann Method for Fluid Flows, \href{https://doi.org/10.1146/annurev.fluid.30.1.329}{Annu. Rev. Fluid Mech., 30:329-64(1998)}. 
\bibitem{jcp} B.N. Todorova and R. Steijl, Quantum algorithm for the collisionless Boltzmann equation. \href{https://doi.org/10.1016/j.jcp.2020.109347}{Journal of Computational Physics 409, (2020)}.
\bibitem{qip}  L. Budinski, Quantum algorithm for the advection-diffusion equation simulated with the lattice Boltzmann method. \href{ https://doi.org/10.1007/s11128-021-02996-3}{Quantum Inf Process 20, 57 (2021)}.
\bibitem{qip2} L. Budinski, Quantum algorithm for the Navier-Stokes equations by using the streamfunction-vorticity formulation and the lattice Boltzmann method. \href{https://www.worldscientific.com/doi/abs/10.1142/S0219749921500398}{International Journal of Quantum Information 20, 2150039 (2022)}.
\bibitem{succi1} W. Itani, and S. Succi, Analysis of Carleman Linearization of Lattice Boltzmann, \href{https://doi.org/10.3390/fluids7010024}{Fluids,  7, 24(2022)}.
\bibitem{qip3} D. Wawrzyniak, J. Winter, S. Schmidt, T. Indinger, C. F. Janßen, U. Schramm, and N. A. Adams,
A quantum algorithm for the lattice-Boltzmann method advection-diffusion equation, \href{https://doi.org/10.1016/j.cpc.2024.109373}{Computer Physics Communications 306, 109373(2024)}.
\bibitem{succi} W. Itani, K. R. Sreenivasan, and S. Succi, Quantumalgorithm for lattice Boltzmann (QALB) simulation of incompressible fluids. \href{https://doi.org/10.1063/5.0176569}{Physics of Fluids 36, 017112 (2024)}.
\bibitem{qip4} D. Wawrzyniak, J. Winter, S. Schmidt, T. Indinger, C. F. Janßen, U. Schramm, and N. A. Adams, Unitary Quantum Algorithm for the Lattice-Boltzmann method.arXiv:2405.13391v3.
\bibitem{qip5}  S. Kocherla, A. Adams, Z. Song, A. Alexeev, and S. H. Bryngelson, A two-circuit approach to reducing quantum resources for the quantum lattice Boltzmann method. arXiv:2401.12248v2.
\bibitem{qip6} S. Succi,W. Itani, K.Sreenivasan, and R. Steijl, Quantum computing for fluids: Where do we stand? \href{https://iopscience.iop.org/article/10.1209/0295-5075/acfdc7}{Europhysics Letters, 144, 10001(2023)}.
\bibitem{FDM}  D. W. Berry, High-order quantum algorithm for solving linear differential equations. \href{doi:10.1088/1751-8113/47/10/105301}{Journal of Physics A: Mathematical and Theoretical,  47: 105301(2014)}.
\bibitem{FDM1}  R. Demirdjian, D. Gunlycke,  C. A. Reynolds,  \textit{et al.}, Variational quantum solutions to the advection–diffusion equation for applications in fluid dynamics.\href{https://doi.org/10.1007/s11128-022-03667-7}{Quantum Information Process 21, 322 (2022)}. 
\bibitem{state} L. Grover and T. Rudolph, Creating superpositions that correspond to efficiently integrable probability distributions. \href{
https://doi.org/10.48550/arXiv.quant-ph/0208112}{arXiv:quant-ph/0208112}.
\bibitem{pra}  Y. Sato, R. Kondo, S. Koide, H. Takamatsu, and N. Imoto,  Variational quantum algorithm based on the minimum potential energy for solving the Poisson equation. \href{https://doi.org/10.1103/PhysRevA.104.052409}{PHYSICAL REVIEW A 104,052409(2021)}.
\bibitem{np} HY. Huang, R. Kueng, and J. Preskill, Predicting many properties of a quantum system from very few measurements. \href{https://doi.org/10.1038/s41567-020-0932-7}{Nature Physics 16, 1050-1057 (2020)}. 


\end{thebibliography}
\end{document}